\documentstyle[epsfig]{mn2e}

\title[Disc inner radius in black hole binaries]
{The soft component and the iron line as signatures of the disc inner radius in Galactic black hole binaries}

\author[Kolehmainen, Done \& D{\'{\i}}az Trigo]
{Mari Kolehmainen$^{1,2}\thanks{E-mail:mari.kolehmainen@astro.ox.ac.uk}$, Chris Done$^2$ and Mar{\'{\i}}a D{\'{\i}}az Trigo$^{3}$ \\
$^1$School of Physics \& Astronomy, University of Southampton, Highfield, Southampton, SO17 1BJ, UK \\
$^2$Department of Physics, University of Durham, South Road, Durham
DH1 3LE,
UK\\
$^3$European Southern Observatory, ALMA Regional Centre, Karl-Schwarzschild-Str. 2, D-85748 Garching, Germany \\
}

\pagerange{\pageref{firstpage}--\pageref{lastpage}} \pubyear{2010}

\begin{document}

\topmargin = -0.5cm

\maketitle

\label{firstpage}

\begin{abstract}

The inner radius of the accretion disc around a black hole in the low/hard state can be measured in 
one of two ways. Firstly, via the extent of broadening of the iron emission line, and secondly, from the 
luminosity and temperature of the weak soft component seen in this  state, assuming it is the disc. We 
use both of these methods on all the low/hard state spectra taken in timing mode of {\sl XMM-Newton}
's EPIC-pn. We find that the two methods are not consistent with each other, and the difference is 
not always in a single direction. The two methods are neither model independent, nor are they 
independent of current calibration issues. We find that the remaining small residuals in the EPIC-pn 
timing mode response at the $\leq 3$\% level can have a dramatic effect on the fit parameters for the 
reflected spectrum. There is also a mismatch in cross-calibration with {\sl RXTE}, which makes it 
difficult to use simultaneous data to extend the bandpass of the spectral fits. Nonetheless, it is clear 
from the data that the iron line is noticeably broader and stronger at higher $L/L_{Edd}$, which is 
consistent with the truncated disc models. We also show that it is likely that the soft 
component changes character, from a stable component consistent with a truncated disc at high $L/L_{Edd}$, to a variable one with 
much smaller radius at low $L/L_{Edd}$. This adds to growing evidence for a complex soft component 
in the low/hard state, possibly resulting from clumps torn from the edge of the truncated disc.

\end{abstract}
\begin{keywords} accretion, accretion discs, black hole physics, relativity, X-rays: binaries
\end{keywords}

\section{Introduction}

The current paradigm for the structure of the accretion flow in black
hole binaries (hereafter BHB) at low luminosities is that the cool,
optically thick, geometrically thin standard accretion disc is
progressively replaced in the inner regions by a hot, optically thin,
geometrically thick flow as the mass accretion rate decreases
(low/hard state, Esin et al. 1997). This model has gained widespread
acceptance by its ability to provide a framework in which to interpret
large amounts of apparently unrelated observational data,
predominantly revealed by the multiple {\sl RXTE} observations of
these systems. At the lowest luminosities, the large disc truncation
radius means that the disc emission is cool and dim. Few seed photons
from the disc illuminate the flow, so the Comptonised spectra are
hard. Decreasing the disc truncation radius leads to a stronger disc
component, and to a greater overlap of the flow with the disc. This
gives more seed photons to Compton cool the flow, giving softer
Compton spectra. The decreasing radius also means that any frequencies
set by this radius will increase, giving a qualitative description of
the increasing characteristic frequencies seen in the power spectra
and their tight correlation with the energy spectra. The flow is
completely replaced by the disc when the disc reaches its minimum
radius of the last stable circular orbit (high/soft state), giving a
physical mechanism for the marked hard-to-soft transition seen in
black hole binaries.  Even the jet behaviour can be tied into this
picture, as a large scale height flow is probably required for jet
formation, so the collapse of the inner flow as the disc reaches its
minimum radius triggers a similar collapse of the radio emission (see
e.g. Fender et al. 2004, and the reviews by McClintock \& Remillard (2006) and Done, Gierli{\'n}ski
\& Kubota (2007), hereafter DGK07).

Despite these evident successes, these models remain controversial due
to reports that the disc extends down to the last stable orbit in the
low/hard state. There are two observational signatures of
this. Firstly, reflection of the Comptonised emission from the disc is
smeared by a combination of special and general relativistic effects,
and the extent of this broadening is determined by the inner disc
radius (e.g. the review by Fabian et al. 2000). Secondly, the
luminosity and temperature of the direct continuum from the disc
itself can be used to evaluate the emitting area, and hence the inner
disc radius. Both these require CCD data rather than the more numerous
proportional counter {\sl RXTE} datasets (lower energy bandpass for
the low temperature disc emission, and higher spectral resolution for
the iron line profile).  A recent review of low/hard state CCD spectra
from BHB by Reis et al. (2010, hereafter R10) noted that both these
signatures were routinely seen at a level which generally excluded a
truncated disc.

These reports are themselves controversial, and have been challenged in literature. The most convincing broad iron line profile in R10 is
from a bright low/hard state of GX~339$-$4. This profile is derived
from data where instrumental pileup is an issue (Miller et al. 2006;
Done \& Diaz Trigo 2010).  Simultaneous data from another instrument
which does not suffer from pileup clearly shows a much narrower line
(Done \& Diaz Trigo 2010). However, simulations of pileup do not
produce an artificially broad line (Miller et al. 2010), but our
understanding of pileup for such an extreme count rate (200$\times$
over the limit for the instrument mode used) is probably not complete
(see also counterexamples in the data compilations of Ng et al. 2010;
Yamada et al. 2011).

The intrinsic disc emission has a different set of issues. Firstly it
can be much weaker than the Compton continuum even in the CCD X-ray
bandpass, so its luminosity and temperature depend on how the
continuum is modelled (e.g. the difference in inner radius in Rykoff
et al. 2007 from using Comptonised emission compared to a power
law). This is unlike the situation in the high/soft state, where the
disc dominates and the high energy continuum model has little effect on
the results (e.g. Kubota \& Done 2004). Even having modelled the disc
emission, its luminosity and temperature need not be simply due to
gravitational energy release as in the high/soft state. X-ray heating
from illumination by the much stronger hard X-ray component can change
the derived inner disc radius from being consistent with the last
stable orbit (Rykoff et al. 2007) to being considerably larger,
especially as the standard stress-free inner boundary condition is
probably not appropriate for a truncated disc (Gierli{\'n}ski, Done \&
Page 2008).

However, it is also possible that the disc is considerably more
complex. Firstly, even disc dominated high/soft spectra are not
completely described by current disc models. They are
broader than a simple sum of blackbodies, as expected due to
relativistic smearing, and fit much better to models which incorporate
this as well as full radiative transfer through the disc
photosphere. While this makes a very nice physical picture, the disc
spectra are even better fit by phenomenological 
models, showing the limitations of the best current theoretical
descriptions of disc spectra (Kolehmainen, Done \& Diaz Trigo
2011, Kubota et al.2010). Secondly, the disc need not be a single structure.  The inner
edge of the truncated disc is not likely to be be smooth.  Clumps
torn off the disc edge will spiral inwards into the hot flow, so will heat
up by thermal conduction and evaporate. Before they completely
merge into the hot flow they can form a small area, hotter, soft
component, separate from the main body (and spectrum) of the truncated
disc itself (see Figures 9 and 10 in Chiang et al. (2010), Yamada et al. 2013).

As well as potential complexity of the disc spectrum, there is also
potential complexity of the Compton continuum.  At low luminosities
the hot flow should be quite optically thin, in which case Compton
scattering gives a bumpy rather than smooth power law spectrum.  At
higher luminosities the flow has higher optical depth, so can be
inhomogeneous, with different parts of the flow giving different
Comptonised spectra. This is required in order to produce the observed
spectral lags, where the soft continuum varies before the harder
continuum (Miyamoto et al 1988; Kotov et al 2001; Arevelo \& Uttley
2006). Even more direct evidence for this is seen in the frequency
resolved spectra, where the most rapidly variable parts of the flow
(few 10s of milliseconds, presumably the inner regions) have harder
spectra and less reflection than the more slowly variable emission
(few seconds, presumably the outer parts of the flow: Revnivtsev et
al. 1999, Axelsson et al 2013). This gives rise to spectral curvature, which can be seen in
broadband data (di Salvo et al 2001; DGK07; Makishima et al. 2008;
Kawabata \& Mineshige 2010; Shidatsu et al 2011; Yamada et al 2013). Fitting such
continuua with a single Comptonisation component leads to a
requirement for an additional soft component, but this is connected to
the Comptonisation region rather than to the disc.

Thus there is controversy both from instrumental effects for these
bright sources (iron line), and over the physical interpretation of
what is seen (origin of the soft X-ray component). We pick one
particular instrument configuration, that of {\sl XMM-Newton} timing
mode, as this is specifically designed to observe bright sources, and
systematically examine all low/hard state spectra taken in this mode
to date. We assess the effects of both instrumental and modelling
uncertainties, and show that the both the iron line and intrinsic disc
emission can be consistent with the truncated disc models in all
current low/hard state spectra.

\section{Observations and data analysis overview}
\label{analysis}

Galactic black hole binaries are generally too bright to be observed
in the standard imaging modes of CCD detectors, even in the low/hard
state. We are therefore restricted to fast timing modes, which are
currently less well calibrated than the the standard imaging modes
usually used for fainter sources. We select the EPIC-pn timing mode of {\sl
XMM-Newton}, as this is the mode which normally maximises the
non-piled up count rate for low/hard state BHB.

There are 7 archival observations of canonical low/hard states from 4 sources in
this mode: Cygnus~X-1, Swift~J1753-0127, GX~339$-$4 (4 datasets) and
H1743-322. The latter object has an interstellar column density of
$\sim 10^{22}$~cm$^{-2}$, substantially higher than the others. This
means that the low energy continuum emission in H1743-322 is much less
visible. This clearly reduces the constraint on the intrinsic disc
emission, but also affects the iron line, as the latter depends on
accurate modelling of the continuum emission underneath the line (see
e.g. Kolehmainen et al.  2011), which in turn requires broad bandpass
data. Thus we exclude H1743-322 from this analysis (see Table~
\ref{obs}). We also considered the single archival observation of a recently discovered black hole candidate XTE~J1752-223 (Markwardt et al. 2009), which caught the source towards the end of a soft-to-hard state transition. However, on closer look the spectral shape of the observation resembles more that of a hard-intermediate state spectrum, which was also confirmed by the hardness-intensity and rms-properties of simultaneous {\sl RXTE} observations. Thus, since this observation is not in the {\sl canonical} low/hard state, it was omitted from our analysis.  

\begin{table}
\begin{tabular}{l|l|l|l|l}
\hline
 &  Obsid & cts/s\footnote{Footnote} & Exp (s)  \\
\hline 
Cygnus~X-1 & 0602610401 & $479\pm0.4$ (1111) & 19970 \\
 GX~339$-$4 (GX4) & 0654130401 & $362\pm0.3$ (944) & 25290 \\
 GX~339$-$4 (GX3) & 0204730301 & $257\pm0.2$  & 44360 \\
 GX~339$-$4 (GX2) & 0204730201 & $240\pm0.2$  & 30480    \\
 GX~339$-$4 (GX1) & 0605610201 & $125\pm0.1$  & 31750    \\
 Swift J1753-0127 & 0311590901 & $85\pm0.1$  & 40110 \\
\hline
\end{tabular}
\caption{Details of the {\sl XMM-Newton} observations analysed in this paper. The highly-absorbed BHB H1743-322 (${\rm N_{H}}\sim16\times10^{21}$) was excluded due to the absorption's obscuring effect at low energies. The binary parameters used in this paper are Cygnus X-1: ${\rm M=20M_{\odot}}$, ${\rm D=24kpc}$, ${\rm i=30^{\circ}}$, GX~339$-$4: ${\rm M=10M_{\odot}}$, ${\rm D=8kpc}$, ${\rm i=60^{\circ}}$ and Swift J1753: ${\rm M=9M_{\odot}}$, ${\rm D=6kpc}$, ${\rm i=60^{\circ}}$. }
\label{obs}
\end{table}

The data were reduced using the \textit{XMM-Newton} Science Analysis
System (SAS) v10.0. We applied the standard data reduction
expressions, using single and double events and ignoring bad
pixels. All data were extracted in full RAWY [1:200] and RAWX of 6
rows on either side of the central row. The SAS tool {\sc epatplot} was
used to check for pile-up in all of the observations. This showed that
Cyg X-1 and the brightest low/hard state observation of GX~339$-$4
were slightly affected. This was corrected by excluding 1 row in RAWX
on both sides of the peak of the emission. Response and ancillary
files were generated with SAS tasks {\sc rmfgen} and {\sc arfgen},
respectively. The spectra were then rebinned using {\sc specgroup},
with an oversampling factor of 3, as recommended for all of the EPIC
detectors. Each bin was also set to have a minimum of 25 counts, and a systematic error of 1\% was added to the spectrum.

The current level of EPIC instrument calibration is discussed at
length in the latest version of the {\sl XMM-Newton} Calibration Technical
Note (0083)\footnote{http://xmm.vilspa.esa.es/external/xmm\_sw\_cal/
\\ calib/documentation.shtml}. The main issue for concern is the
ubiquitous problem of X-ray loading (XRL) in observations taken before May 2012. The 'quiet' level of the
electron current in each pixel is determined from exposures at the
beginning of each observation, and this offset map is automatically
subtracted from the data by the onboard processor. However, for bright
sources, and especially bright, hard sources, this electron current is
contaminated by the source itself. The source pixels have too much
electron current subtracted, leading to an offset in the gain for each observation. 
However, as of 30th May 2012, the filter is closed at the start of each observation 
and the effect is not present in observations after this date. The newly published SAS13.0 also includes 
a task for correcting for the effects of XRL in the standard imaging mode, but not the fast timing modes.

The charge transfer inefficiency (CTI) is a separate issue, and does
not include the effects of XRL. Damage to the CCD mean that there are
electron traps, so not all charge is transferred on readout. This
charge transfer inefficiency is reduced for bright sources, as the
multiple electrons produced by high X-ray illumination
fill the holes, so that the remaining charge can be
efficiently transferred. This is currently corrected by the SAS task {\sc epfast}, but
the parameters for the gain shift were derived assuming that the data
were affected only by a linear gain shift, while in reality they are
affected by a combination of a linear gain shift from CTI and a
constant offset from XRL. We follow current
recommendations and use {\sc epfast} on all our data, but at these 
relatively low count rates (compared to the ones seen in burst mode)
the correction did not cause any noticeable changes in the data.

The wings of the point-spread function of the EPIC-pn extend further
than the data collection region in timing mode, which means that
selecting a source-free region for background subtraction is not
possible (e.g. Done \& Diaz Trigo 2010). However, the 10--15~keV
light curve from the outer regions of the EPIC-pn can still be used to
identify and exclude periods of background flaring, and these can be
checked from outer chip light curves from the MOS imaging data, when
available. We also use blank sky backgrounds in timing mode to check
that the background is indeed negligible for these bright, hard sources.
Table 1 shows the resulting effective exposure time after excluding
period of background flaring.

We also extract the quasi-simultaneous {\sl RXTE} data on all of our
objects and reduce these using standard methods. Table~\ref{obs} gives
details of the observations used.
                                                                                                             
\section{low/hard spectra overview}
\label{continuum}

\begin{figure}
\begin{center}
\leavevmode\epsfig{file=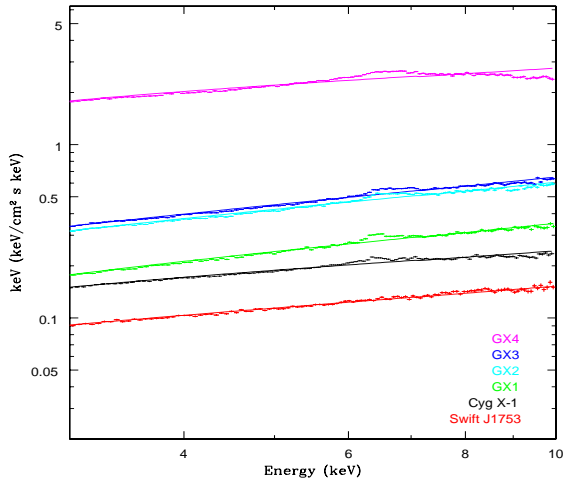,height=7.0cm,width=8cm}
\end{center}
\caption{All the observations analysed in this paper, unfolded by a simple power law model to illustrate the spectral deviations from a simple continuum. The spectra are plotted in order of increasing luminosity in the 3--10~keV range.}
\label{all_data}
\end{figure}

\begin{figure}
\begin{center}
\leavevmode\epsfig{file=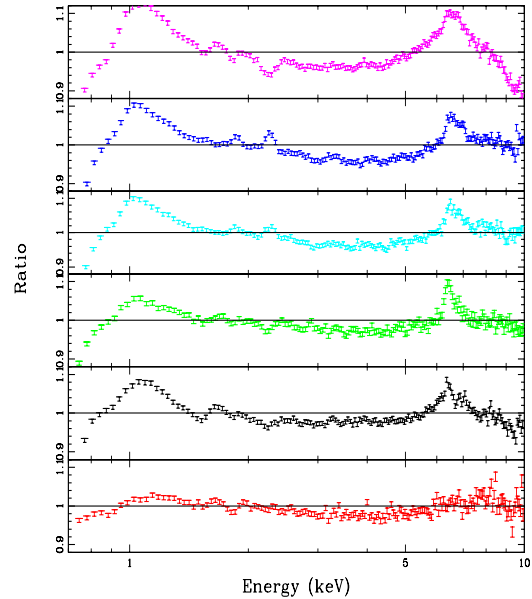,height=9.0cm,width=7.5cm}
\end{center}
\caption{The data/model ratio of all the observations using absorbed power law in the 0.7--10~keV range., with the same colour convention as in Figure~\ref{all_data}. The data plotted in order of increasing luminosity to illustrate the changes in the soft excess and the iron line.}
\label{all_softex}
\end{figure}

We start by fitting the data with a single power law model to illustrate any possible deviations from a pure 
power law continuum. Figure~\ref{all_data} shows the data unfolded with this model and fitted in the 3--10~
keV band. For plotting purposes we have divided Cyg X-1 and Swift J1753-0127 (hereafter S1753) by $(M/
10)(8/D)^2$ so that the spectra are roughly normalised in relative $L/L_{Edd}$ assuming that GX~339$-$4
is a 10M$_{\odot}$ black hole at 8kpc. 

The data show a clear trend in luminosity. At the lowest luminosities (S1753 in Figure~\ref{all_data}) the 
spectrum shows no clear features or deviations from a simple power law continuum. At higher luminosities the 
excess emission around the iron line region increases in both strength and width with increasing $L/L_{Edd}$. 
This effect is even more pronounced in Figure~\ref{all_softex}, where the data are modelled with an 
absorbed power law in the 0.7--10~keV range and plotted as a data/model ratio. The hydrogen column 
density is let to vary within reasonable limits. The increasing soft excess at $\sim$1~keV is now also clearly 
visible. Assuming that this soft excess is the true disc component, these features are qualitatively 
consistent with the predictions of the standard truncated disc model. At the lowest luminosities the inner disc 
is truncated far from the last stable orbit, then subsequently  reaching further 
inwards and becoming stronger as the 
source gets brighter. The reflection fraction also increases with luminosity, which could at least partially 
explain the change in the continuum shape above the broad iron line region in the brightest observation (GX4)
. We explore different explanations for this in Section~\ref{s1753} and further in the paper.

\subsection{Cross-Calibration with RXTE}

\begin{figure}
\begin{center}
\leavevmode\epsfig{file=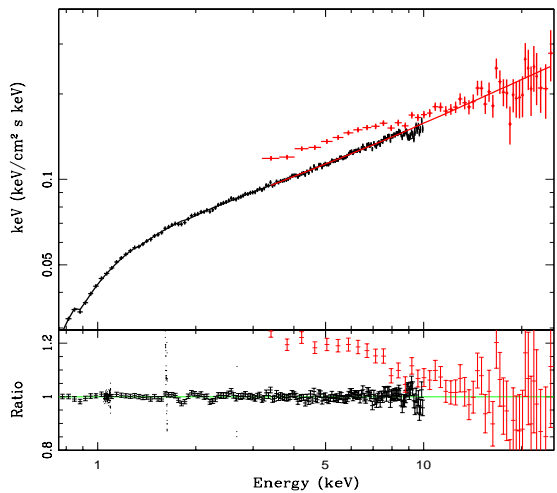,height=7.0cm,width=8cm}
\end{center}
\caption{Joint EPIC-pn-PCA fit of the S1753 observation fitted in the joint energy range of 0.7--25~keV. The data are consistent only above 7~keV, with the difference in photon indices of $\Delta\Gamma\sim0.11$. Due to this obvious disagreement in cross-correlation, the rest of this analysis focuses  solely on the EPIC-pn data. }
\label{pnpca}
\end{figure}

As the EPIC-pn energy band only reaches up to 10~keV, we initially combine it with data 
from the {\sl RXTE} PCA to cover more of the hard X-ray tail. However, fitting data from the two instruments showed an inconsistency in their cross-calibration. We demonstrate this with S1753, which has the simplest spectrum of our sample, with very little spectral features or curvature (see Figures~\ref{all_data} \& \ref{all_softex}). Fig~\ref{pnpca} shows the simultaneous EPIC-pn/PCA data of the same source.  This shows a clear discrepancy in the cross-calibration of the two
instruments in the region of overlap (3-10~keV). This was also noted
in Hiemstra et al. (2011) for the bright BHB XTE J1652. However, for
S1753 there is almost no spectral complexity to mask the issues.
The two instruments clearly have different spectral indices, with
$\Delta\Gamma=0.11^{+0.01}_{-0.02}$, even restricting the fit to the
3-10~keV region where the data overlap. We find similar discrepancies
in spectral indices in all our data in the overlapping 3-10~keV
bandpass, though here the evident complexity around the iron line
could affect the modelling. This issue was also noted in the latest update to the {\sl XMM-Newton} Calibration Technical
Note (v.1.6 of TN-0083), where the comparison was made between 
{\sl RXTE}/PCA and the EPIC-pn burst mode, rather than timing mode (see also the Appendix). 
Even though the EPIC-pn science modes are different, the discrepancy is 
in the same direction, i.e. the PCA spectra are softer than 
the EPIC-pn. 

Thus it is clear that these data are still somewhat limited by 
calibration, and understanding these limitations and their origin is 
essential before making interpretations of such data. This offset in 
spectral index means that we cannot fit the EPIC-pn and the PCA data 
together. This is a general problem with cross-calibration between the 
two detectors, rather than a feature of timing mode alone 
(see the Appendix for the same issue in 
Burst and Imaging mode). 
Hence we focus the rest of this analysis solely on the EPIC-pn 
data, as we are primarily interested in the soft continuum and the iron 
line profile.

\section{Lowest ${\rm L/L_{Edd}}$: Swift J1753-0127}
\label{s1753}

We start the more detailed analysis with the simplest spectrum, S1753. 
Previous work on this spectrum has 
shown that there is
curvature in the continuum, which can be described either by a disc
component or by reflection (R10, Hiemstra et al. 2011).  However, both
of these would imply that the inner disc is present at some level, in
conflict with the simplest truncated disc model. 

We confirm that there is indeed spectral curvature by fitting a series
of models of increasing complexity. We start with a single 
Comptonisation component, described by the {\sc nthcomp} model of
Zdziarski et al. (1996), as a power law is a poor approximation for Comptonisation
where the bandpass is close to the seed photons. We assume these seed photons
have a blackbody shape, and fix the electron temperature at 100~keV. 
We absorb this continuum so the 
total model is {\sc tbabs*nthcomp}, giving 
$\chi^2_\nu=196/163$ for a seed photon temperature of $\sim 0.2$~keV. We
add a disc spectrum with inner disc temperature tied to the
seed photon temperature for Comptonisation i.e. {\sc tbabs*(diskbb+nthcomp)}. This gives a significantly
better fit with $\chi^2=169/162$. The disc normalisation of
$400_{-100}^{+180}$ implies an apparent radius of 17~km for the fiducial
values of distance and inclination.  This gives a corrected radius of
20~km for a colour correction factor of 1.7 and stress free inner
boundary condition of 0.41 (Kubota et al. 2001), which is $1.3R_g$ for
the fiducial mass of $10M_{\odot}$. Even without the stress free inner
boundary the radius is only $3R_g$, so this is completely inconsistent
with a truncated disc. Instead, this small emitting area could 
be indicative of small clumps at large radii torn from the truncated
disc edge, heated by conduction as they spiral into the hot flow
(Chiang et al. 2010).

Instead, we get an even better fit using the {\sc eqpair}
Comptonisation model with no additional soft component
($\chi^2_\nu=160/163$). This model calculates the full Comptonised emission
from each individual Compton scattering order, so at low optical
depths and high temperatures (best fit is $\tau\sim 0.3$,
$kT_e=300$~keV) the excess soft X-ray flux is fit by the first order
scattering from seed photons from the disc at $11\pm 1$~eV. The disc
normalisation is completely unconstrained at these low temperatures,
so is consistent with a truncated disc.

Thus the {\em spectrum} of S1753 is consistent within current
instrumental uncertainties as being simply described by a single
Comptonisation continuum from low optical depth material, with no disc
required in either direct or reflected emission, as predicted by the
truncated disc models at low $L/L_{Edd}$. However, this is not
consistent with the {\em timing} behaviour.  Fast time variability shows
clearly that there {\em is} an additional component at soft energies
(Figure 3 in Uttley et al. 2011). This soft component leads the harder X-rays
by $\sim 0.1$~s, far too long to be the light crossing time lags
between individual Compton scattering orders (see also Miyamoto \&
Kitamoto 1988). Instead, these almost certainly are viscous lags from
propagating fluctuations, with fluctuations in a spectrally softer
component at larger radii propagating down to modulate a harder
component produced at smaller radii (Kotov et al 2001; Arevalo \&
Uttley 2006).

Hence we do want to include a separate soft component in the {\sl XMM-Newton}
bandpass in these data. If this is roughly blackbody in shape then it
could either represent the inner edge of an untruncated disc around an
extreme spin black hole, or small clumps torn from the edge of a disc
which is truncated at much larger radii. Clumps have the advantage of 
also giving a clear origin for variability, whereas a disc down to the
last  stable orbit in the disc dominated states has remarkably little
variability (e.g. Churazov et al. 2001). 

\begin{figure}
\begin{center}
\leavevmode\epsfig{file=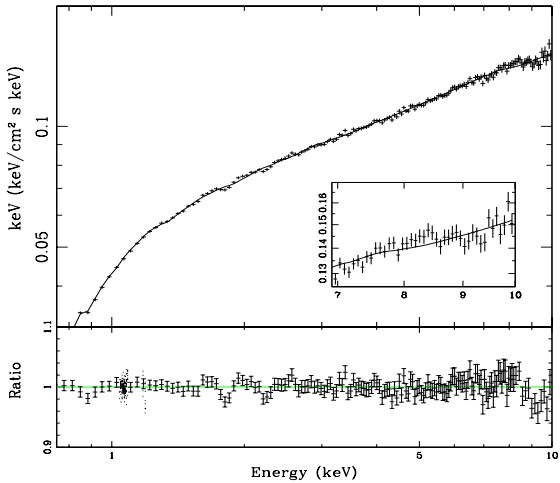,height=7.0cm,width=8cm}
\end{center}
\caption{The S1753 observation modelled with a simple {\sc diskbb+nthcomp} continuum plus reflection, and zoomed in to the 7--10~keV region. A $\sim6$ per cent dip is visible in the residuals at $\sim$9~keV.}
\label{nonotch}
\end{figure}

Observationally, these two possibilities predict different reflection
signatures. Small clumps subtend very little solid angle, so
give a small reflected spectrum which is not strongly
smeared. Conversely, an inner disc round a high spin black hole should
be physically close to the X-ray source, so should give a larger
reflected fraction and strong relativistic smearing.  We include
reflection of the Comptonisation continuum from 
ionised material, modelled using the {\sc rfxconv} model, 
based on the tables of Ross \& Fabian (2005) recoded as a
convolution model (Kolehmainen, Done \& Diaz Trigo 2011). This is 
relativistically smeared using the Kerr metric transfer functions of 
Laor (1991), recoded as a convolution model. Thus the total model is
{\sc
tbabs*(diskbb+nthcomp+kdblur*rfxconv*nthcomp)}, and this 
gives
$\chi^2_\nu=140/159$, $\Delta\chi^2=25$ for three additional
parameters better than the original {\sc diskbb+nthcomp} continuum
model.  The disc normalisation is now allowed to be much larger, up
to $2400$, but this does not make the model compatible with a truncated disc as the 
reflected emission requires strong relativistic smearing,
with $r_{in}=1.9^{+6.9}_{-0.6}R_{g}$ even though the amount of
reflection is small ($\Omega/2\pi = 0.06^{+0.04}_{-0.02}$). The reason
that the data require such small radii is that the drop from the blue
wing of the iron line is at $\sim 8.5$~keV (Fig.~\ref{nonotch}), so
requiring large Doppler blueshifting from the rest line energy for
He-like iron (as implied by the ionisation state) of 6.7~keV.

At first sight this strongly supports the untruncated disc. However,
the parameters are puzzling in this geometry. The amount of reflection
is very small, which, together with the hard continuum, supports
models where the X-rays are beamed away from the disc (e.g. Malzac,
Beloborodov \& Poutanen 2001). However, this also changes the illuminating
radiation pattern, defocusing it away from the disc central regions. Yet
the reflection spectrum requires that the inner disc is illuminated in
order to produce the observed smearing.  This, together with the fact
that the features being fit by reflection are very small (less than a
few percent in a ratio plot) means that they are critically dependent
on the current calibration of the {\sl XMM-Newton} EPIC-pn timing mode. We
explore this in more detail below by using a combination of all the 
low/hard state spectra, and the Crab data. 

\section{All low/hard state spectra and limitations of the current EPIC-pn
  timing mode response}

\begin{figure}
\begin{center}
\leavevmode\epsfig{file=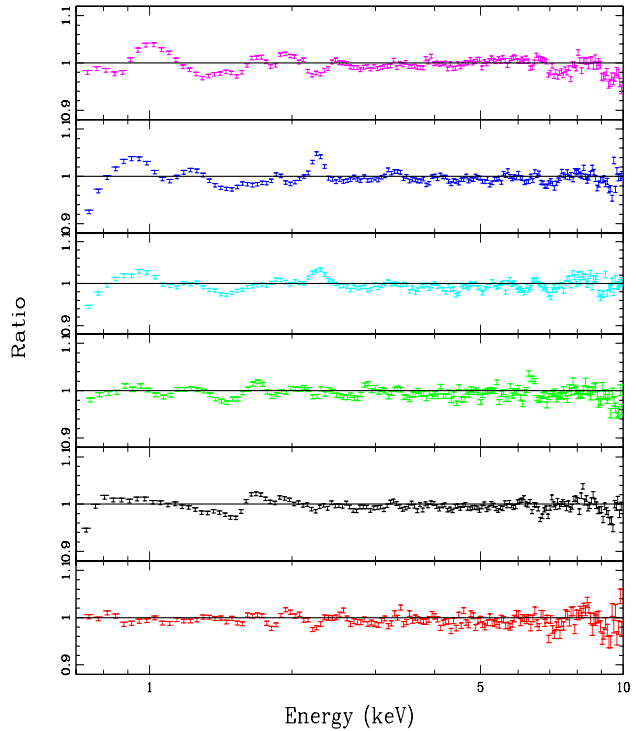,height=11cm,width=9cm}
\end{center}
\caption{Data/model ratio for all spectra, when fitted with {\sc diskbb+nthcomp+reflection} model. }
\label{ratio_all}
\end{figure}

We fit all the spectra with a continuum model of a disc plus single 
Comptonisation model
i.e. assume that there is a real additional soft component. This is supported
by the fact that Cyg X-1 and GX1,2,3 all have similar lag spectra to
S1753 (GX4 is too recent an observation to be included in Uttley
et al. 2011).  Figures~\ref{all_data}\&\ref{all_softex} show that all the 
spectra, apart from S1753, have obvious residuals from reflection, so we 
also include reflection in the model {\sc tbabs*(diskbb+nthcomp+
kdblur*rfxconv*nthcomp)}. Both GX~339$-$4 and
Cyg X-1 also require a small, narrow, neutral core to the iron line,
which is probably due to reflection from the raised rim of the outer
disc. We include this as a narrow ($\sigma$ fixed to 0.01~keV)
neutral line (energy fixed at 6.4~keV) in the model.  All our fit
parameters are shown in Table~\ref{diskbb} and residuals are plotted
in Figure~\ref{ratio_all}. 

The residuals to the best fit models clearly show increasing problems
around the edges in the response matrix ($\sim$1.8 and $\sim$2.2~keV)
as the source count rate increases, presumably due to the uncorrected
effect of XRL.  However, spectrum GX3, taken 1 day after spectrum GX2
(consecutive {\sl XMM-Newton} orbits) and with very similar count rate 
and
spectral shape, has stronger residuals at 2.2~keV, indicating a
different gain shift. This cannot be due to different XRL as the count
rate is very similar. Instead it most likely shows the systematic 
uncertainty of the EPIC-pn timing mode calibration. The stronger residuals 
around the instrument
edges affect some of the fit parameters, so that the disc
normalisation changes significantly. This is almost certainly an
artefact, as real changes in geometry are not likely to occur on such
short timescales without a correspondingly change in flux.  Hence it
seems most likely that the difference between GX2 and GX3 is driven
mainly by unknown, time dependent stability issues in the EPIC-pn
response.

There is also an excess at 1~keV, which appears systematically
stronger at higher $L/L_{Edd}$.  Such an excess is often seen in
heavily absorbed systems (e.g. Heimstra et al. 2011), where it may be
a symptom of the uncertainties in the low energy tail of the response
to higher energy photons. However, this
should not be an issue for the low-to-moderate absorption columns
required here. Thus it is likely to be real.
Nonetheless, it is not
easy to interpret physically, despite the energy pointing to iron L
emission, as the reflector required to make the iron K line is too
highly ionised to produce much iron L.
A reflection origin would also
impact on the lag spectrum, turning the generic hard lag into a lead
at these energies as the reflected emission will follow the hard X-ray
illumination (e.g. Madej et al. 2012). The lag spectra show no such
feature (Uttley et al. 2011). Hence this most probably shows that the
soft component is not well modelled by {\sc diskbb}, but instead has a
more complex spectrum (see also Shidatsu et al 2013).

There is also a drop above 9~keV which is always present. This feature
could be real if there are substantial amounts of ionised H-like iron
as this has a K-edge energy of 9.28~keV (e.g. Hiemstra et
al. 2011). However, it would then be expected to vary with the amount
of ionised reflection, yet this drop has the same $\sim$5\% level
irrespective of $L/L_{Edd}$. One possible explanation for this edge-like 
feature could be the lack of background subtraction at these high 
energies, rather than an intrinsic
feature in the spectra. However, extracting a background in RAWX [3:10] 
did not make this feature disappear. The presence of a warm absorber 
could also result in a broad edge at these energies, as shown in e.g. Diaz 
Trigo et al. (2012). However, the fact that the same feature seems 
to be present in one of the on-axis EPIC-pn Crab observations (see the 
Appendix), suggests that it might be an instrumental effect. 

\begin{figure}
\begin{center}
\leavevmode\epsfig{file=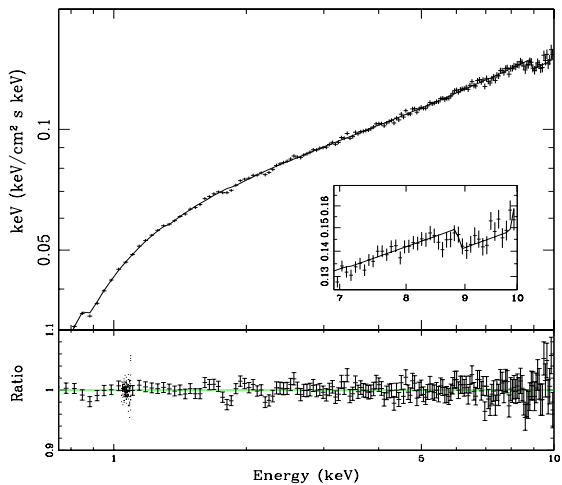,height=7.0cm,width=8cm}
\end{center}
\caption{The S1753 observation modelled as in Figure~\ref{nonotch}, but with a broad absorption line fixed at 9.39~keV as described in Section~\ref{continuum}.   }
\label{swift_notch}
\end{figure}

We revisit all our spectral fits with this caveat. For S1753, the drop
at high energies in the data was the key feature which meant that
highly smeared reflection was significantly detected. With the {\sc notch} 
(Figure~\ref{swift_notch}), the reflection
component is now only marginally significant ($\Delta \chi^2=12$ for 3
additional degrees of freedom), 
though all the model parameters are similar within the
uncertainties. This includes the inner radius, which is still small,
showing that the best fit model has reflection which is strongly smeared,
but the driver for this is now the soft excess at low energies
produced by ionisation reflection rather than the iron line
region. We caution that there is a 30\% difference 
in the amount of low energy reflection between the Ross \& Fabian (2005) 
calculations and the new models of 
Garcia et al (2013). Both codes calculate constant density 
illumination so should be directly comparable, and give very similar
results for softer spectra (Garcia et al 2013).

\begin{figure}
\begin{center}
\leavevmode\epsfig{file=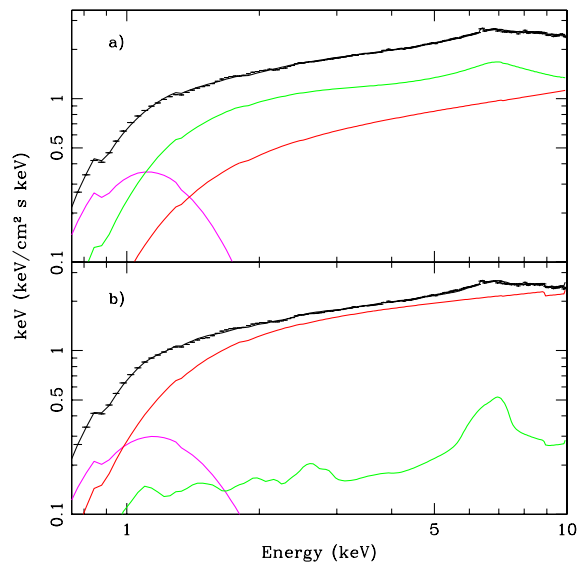,height=8.0cm,width=8cm}
\end{center}
\caption{The spectral decomposition of GX4 without the {\sc notch} (top panel) and with the {\sc notch} (bottom panel). These two very different solutions to the data give very similar ${\chi}^2$ values: 392/158 d.o.f (without notch) and 397/158 d.o.f (with notch). The green line describes the reflection component, the continuum is plotted in red and the soft component in magenta. }
\label{gx4_notch}
\end{figure}

None of the fits to GX1-3 and Cyg X-1 are significantly changed by
inclusion of the {\sc notch}, as reflection is much more significantly
detected in these datasets, and is less strongly smeared in the GX~339$-
$4 observations 
($r_{in,ref}>100$) than in S1753. This makes it much less
dependent on the high energy region, so the reflection parameters are
robust to small changes in effective area at 9--10~keV.  

However, for
GX4, the much broader reflection features mean that the high energy
calibration again makes a difference. Without the {\sc notch}, the 
amount of
reflection is larger than expected for isotropic illumination, with
$\Omega/2\pi=1.3$. The spectral broadening of the iron features which
is also evident in Figures~\ref{all_data}\&\ref{all_softex} is driven mainly 
by its higher ionisation
state, so the derived inner radius is surprisingly large, with
$r_{in,ref}>100$. With the {\sc notch} the amount of reflection drops to
$\Omega/2\pi=0.35$, its ionisation is similar to that in GX1,2,3 and
the obvious broadening is now due to a smaller inner radius, with
$r_{in,ref}=50$. These very different spectral decompositions are
shown in Fig \ref{gx4_notch} a and b. 

We also note that even without the {\sc notch}, the spectral decomposition
in Fig \ref{gx4_notch}a where reflection dominates the 
continnuum in the 1--4~keV bandpass 
is dependent on the detailed shape of the
reflected continuum in this energy range as well as on the shape
of the iron line, so will also be sensitive to the reflection 
model uncertainties discussed in Garcia et al. (2013).
Thus the spectral decomposition in Fig \ref{gx4_notch}a is not robust to both 
current instrumental and model uncertainties.


\section{Disc, double Comptonisation and reflection}

The spectral lags clearly show that there is a separate soft component
in S1753, GX1,2,3 and Cyg X-1 (Uttley et al 2011). While GX4 was
not included in that study, it is plain from the spectrum alone that
there is a separate soft component in these data also, as
$L_{soft}/L_{tot}$ is much larger in this dataset than in the others.
However, the spectral lags are not confined to the soft component
alone. It has long been clear that there is a complex pattern of hard
lags in Compton continuum, which requires an inhomogeneous emission
region. The most successful model to date can match these observed
lags by fluctuations propagating down through the accretion flow,
where the outer parts of the flow have a softer spectrum than the
inner. Two Comptonisation components (together with the disc and
reflected emission) are also required to adequately model the low/hard
state spectra of BHB (di Salvo et al. 2001; Makishima et al. 2008; Yamada et al 2013).  
We describe this additional Comptonisation with the {\sc comptt} model
rather than using another {\sc nthcomp} component so that we can more easily keep
track of each component. 
However, the more limited bandpass of our data mean that we cannot
constrain all the parameters, so we fix the electron temperature of
this additional component at 10~keV.  We assume that both soft and
hard Compton components have the same seed photon energy, and that
this represents the temperature of the disc itself.

An additional soft continuum 
component generically means that the disc component
goes down to lower temperatures, and its normalisation increases, as
does the interstellar column density. This is a nice feature of this
additional component, as all the columns derived from the previous
fits are somewhat lower than expected, except for GX4. However, again
our data cannot constrain all of the parameters, so we fix $N_H$ at
the expected value for all of our data ($0.21$, $0.55$ and $0.55\times
10^{22}$ in S1753, GX~339$-$4 and Cyg X-1, respectively).  We include the
{\sc notch} with parameters fixed to those of the Crab, and tabulate results
in Table~\ref{comptt}.

This model gives significantly better fits than the single Compton
continuum model (compare with Table~\ref{diskbb}) for all spectra
except for GX4. In GX4 the observed, dominant soft component has a
shape which is very similar to a disc. By contrast, in all the other
spectra where the soft component makes only a small contribution to
the spectrum below 1~keV, the shape of this soft component is much
better described by thermal emission plus a broader
spectrum. Conversely, in S1753, the combination of this broader soft
emission plus the {\sc notch} means that reflection is not significantly
detected.

\begin{table*}
\begin{tabular}{lc|l|l|l|l|l|l|l|l|l|l|l|c}
  \hline
\multicolumn{8}{|c|} {\sc TBabs(diskbb+nthComp+Gaussian+kdblur$\times$rfxconv$\times$nthComp)}  \\
 \hline
   \small & $N_H$($\times 10^{22}$) & $T_{in}(keV) $ &
   $N_{Disc}$($\times10^3$) & $\Gamma$  & $N_{comp}$ & $R_{in}$ ($R_{g}$) &
   $f=\Omega/2\pi$ & log$\xi$ & eW (eV) & $\chi^{2}$/ d.o.f   \\ 
\hline
   \hline
 Cyg~X-1 & $4.8\pm0.02$ & $0.21\pm 0.01$ & $170^{+40}_{-30}$ & $1.69\pm0.01$ & 1.48 & $16\pm 4$ & $0.07\pm0.01$ & $2.80^{+0.11}_{-0.05}$ & $8\pm 2$ & 294/158  \\
 GX4 & $6.6\pm 0.2$ & $ 0.18\pm 0.01 $& $660^{+100}_{-80}$ & $1.58\pm 0.02 $ & $0.27$ & $2.2^{*}_{-0.5}$ & $3.54\pm 0.02$ & $270^{*}_{-150}$ & $8\pm 3$ & 381/158\\ 
 GX3 & $3.7\pm0.01$& $0.31\pm 0.01$& $1.6^{+0.4}_{-0.2}$ & $1.55\pm 0.01$ & 0.14 & $92^{+35}_{-20}$ & $0.15^{+0.01}_{-0.02}$ & $2.69^{+0.01}_{-0.11}$ & $5\pm 3$ & 552/158 \\  
 GX2 & $3.9 \pm 0.2$ & $0.27\pm 0.02$& $2.6^{+0.9}_{-0.7}$ & $1.56\pm 0.01$ & $0.14$ & $85_{-25}^{+40}$ & $0.17$ & $2.53^{+0.17}_{-0.08}$ & $5\pm3$ & 358/158	 \\ 
 GX1 & $4.4\pm 0.2$& $0.24^{+0.01}_{-0.02}$& $2.6^{+1.2}_{-0.8} $ & $1.53\pm0.01$ & 9.0 & $110^{+80}_{-40}$ & $0.17^{+0.04}_{-0.03}$ & $2.37^{+0.05}_{-0.02}$ & $13\pm 3$ & 154/158  \\
S1753 & $1.2^{+0.4}_{-0.2}$ & $0.20_{-0.05}^{+0.10}$ & $0^{+2.4}_{*}$ & $1.60\pm 0.01$ & $0.05$ & $1.9^{+6.9}_{-0.6}$ & $0.06_{-0.02}^{+0.04}$ & $2.76^{+0.21}_{-0.05}$ & & 140/159 \\
\hline
\end{tabular}
\caption{Best-fit parameters from the fits with diskbb and a single Comptonisation model, assuming there is a real soft component, representative of an accretion disc. The data/model ratios are plotted in Figure~\ref{ratio_all}. }
\label{diskbb}
\end{table*}

\begin{table*} 
\begin{tabular}{lc|l|l|l|l|l|l|l|l|l|c} 
  \hline
\multicolumn{8}{|c|}{\sc notch$\times$TBabs(diskbb+comptt+nthComp+Gaussian+kdblur$\times$rfxconv$\times$nthComp)} \\
  \hline
   & $T_{in}$(kev) & $N_{Disc}$($\times10^3$) & $\tau$ & $N_{comp}$ & $\Gamma$ & $N_{nthcomp}$ & $R_{in}$($R_{g}$) & $f=\Omega/2\pi$ & log$\xi$  & $\chi^{2}$/ d.o.f
 \\
 \hline
 \hline
Cyg X-1 & $0.17\pm 0.01$ & $490\pm 6$ & $2.3^{+0.1}_{-0.2}$ & 0.32 & $1.40^{+0.03}_{*}$ & 0.56 & $4.5_{-0.5}^{+0.6}$ & $0.15_{-0.02}^{+0.04}$ & $2.72\pm 0.02$ & 222/157  \\
GX4 & $0.21\pm 0.01$  & $140\pm 20$ & * & * & $1.73\pm 0.01$ & 1.1 & $47^{+10}_{-7}$ & $0.35^\pm 0.05$ & $2.44\pm 0.04$ &  381/158 \\ 
GX3 & $0.16\pm 0.01$ & $71^{+12}_{-9}$ & $1.5\pm 0.1$ & 0.044 & $1.40^{+0.01}_{*}$ & 0.13 & $140^{+60}_{-50}$ & $0.18\pm 0.03$ & $2.49^{+0.18}_{-0.04}$ & 360/150 \\
GX2 & $0.16\pm 0.01$ & $65\pm 9$ & $1.6\pm 0.1$ & 0.036 & $1.40^{+0.01}_{*}$& 0.12 & $115^{+85}_{-35}$ & $0.19\pm 0.03$ & $2.45^{+0.07}_{-0.04}$ & 212/157  \\
GX1 & $0.17\pm0.01$ & $26_{-3}^{+5}$ & $1.8\pm 0.2$ & 0.014 & $1.40^{+0.03}_{*}$ & 
0.07 & $150^{*}_{-50}$ & $0.17^{+0.04}_{-0.03}$ & $2.35\pm0.03$ & 150/157  \\
\hline
S1753 & $0.15^{+0.01}_{-0.04}$ & $6.7^{+3.4}_{-1.0}$ & $2.2\pm 0.8$ & 0.011 & $1.40^{+0.1}_{*}$ & 0.03 & * & *  & *& 141/161  \\
\hline
\end{tabular}
\caption{Best-fit parameters from a double Comptonisation model, including the {\sc notch} component. This additional component, together with a broader soft emission, means reflection is not significantly detected in S1753.}
\label{comptt}   
\end{table*}

\section{The changing disc inner radius}

\begin{figure}
\begin{center}
\leavevmode\epsfig{file=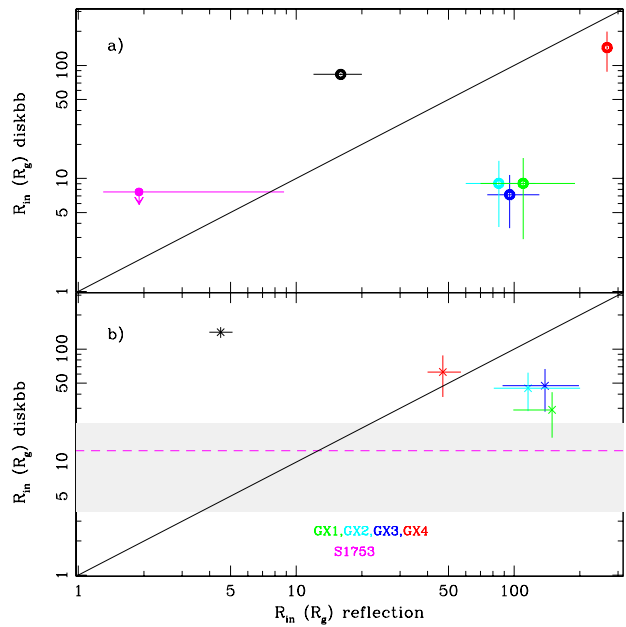,height=9cm,width=9cm}
\end{center}
\caption{The relation between the disc inner radii derived from the disc normalisation and reflection. The line illustrates where the points would lie if these two methods gave consistent results. {\sl a)} The inner radius derived from the soft component (Table~\ref{diskbb}). The S1753 disc normalisation gives only an upper limit for $R_{in}$ when measured from the soft component. {\sl b)} The inner radius derived from the reflection component (Table~\ref{comptt}). Since reflection is not significantly detected in S1753, the dashed line shows the radius based on the disc normalisation, with the grey area illustrating the error in y-direction. GX4 inner radius is the only one in our sample to show consistency within the errors.}
\label{rin}
\end{figure}

Figure~\ref{rin}a  shows the relation between the disc inner radius derived 
from reflection and from the soft component for the single Compton continuum
models (Table 2) while Figure~\ref{rin}b shows this for the double Compton fits. 
The truncated disc model predicts that the disc inner radius decreases
with increasing $L/L_{Edd}$ i.e. should become progressively smaller 
from S1753 (magenta), GX1 (green), Cyg X-1 (black), GX2-3 (cyan/blue), to GX4 (red). 
However, the data do not show this for either model. 

With a single Compton continuum, the dimmest source requires the smallest disc in both soft
emission and reflection (S1753), and the reflection radius strongly requires an untruncated disc. 
GX1-3 are consistent with an untruncated disc from their soft components, while their 
reflected emission indicates a truncated disc. By contrast, in Cyg X-1 both reflection and soft emission 
indicate that the disc is moderately truncated, though the derived radii are not consistent with each other. 
However, the soft component radius is dependent
on the system parameters and for GX~339$-$4 we can compare the disc emission in these low/hard state
data with that derived from  the disc dominated state. We choose the brightest, most disc dominated
observation from those of Kolehmainen et al. (2011) i.e. 
Obsid 0093562701 (burst mode) and derive a {\sc diskbb} normalisation of 
$2.9\pm 0.1\times 10^3$.
This clearly indicates the radius of the 
last stable orbit, and is consistent with the disk normalisation seen in the GX1--2
low/hard state spectra. The disc normalisation then increases quite markedly for the brighter GX4 data. 
If these low/hard state radii derived from the soft component are reliable 
the disc is 
not truncated for the lowest luminosity spectra, but is truncated for 
the brighter low/hard state of GX4,
and then goes back down to the last stable orbit when the source
makes a transition to the disc dominated state. Similar
behaviour is also seen in XTE J1817-330 (Gierli{\'n}ski, Done \& Page 2008), 
where the constant radius soft component seen in the disc dominated
state first increases in radius during the transition to the low/hard state
and then decreases in the dimmer low/hard state. Gierli{\'n}ski et al. (2008)
show that this behaviour can be consistent with the disc progressively truncating 
in the low/hard state as the radii derived for these data are model 
dependent and could be increased by irradiation, and/or a stressed boundary
condition and/or an increased 
colour temperature correction. 
We also note that the disc inner radius is also
increased if the Compton cloud is between the disc and observer, as 
photons in the Comptonised spectrum came originally
from scattering of seed photons from the disc (Kubota \& Done 2004). 

Figure ~\ref{rin}b with radii derived from reflection and emission from the disc with a double Compton
continuum (plus high energy notch for calibration) 
shows a rather different pattern. However, it is still inconsistent
with the overall decrease in radii for brighter low/hard states predicted for the truncated
disc models. The soft component in S1753 still indicates a rather small radius for the disc,
smaller than the much brighter GX4 dataset. However, GX1--4 now do show a marginal
trend of decreasing radius from reflection with increasing $L/L_{Edd}$, opposite to the
increasing radius seen from their soft component normalisation. However, both disc reflection and
emission require that the radius is much larger than the innermost stable circular orbit. 
However, in Cyg X-1, the disc reflection now requires an untruncated disc, while the 
soft component requires a much larger radius.

We note that both Cyg X-1 and GX4 spectra were derived using central column removal. The 
Appendix shows that large changes around the iron line can be 
produced by this process using different energy dependent point spread functions. 
Thus the relativistic 
smearing parameters derived from Cyg X-1 (the most piled-up spectrum used in our analysis) 
are particularly suspect, while GX4 may be rather less affected. 
We note that issues with the 
iron line method were also recently discussed in Dauser et al. 2013, 
and Sanna et al. 2012. In the latter paper 
the line profile implied a physically impossible inner radius of 
$2.0^{+0.4}_{-0.2}R_{g}$ for the neutron star 4U 1636$-$53 (Table 1, 
Sanna et al. 2012).

Larger features are clearly more robust than small, so even though GX4 has central colums removed, this
is not going to remove the large soft component which is clearly detected. Also, by eye, the amount of 
reflection shows a clear trend as a function of $L/L_{Edd}$ (Fig 1). At the
lowest luminosities the iron line is not required, then reflection
becomes a significant feature in the data, though with small solid
angle $\Omega/2\pi\sim 0.2$. This reflecting material is 
significantly ionised. For bright low/hard states (GX4) reflection is 
much stronger, and much more obviously broadened.
This is all consistent with the truncated disc models. What is not, however, is the smaller 
radius inferred for the soft component in the lower luminosity spectra GX1-2 and S1753,
though we note that this still implies a truncated disc in the double Compton continuum models. 
Instead, we suggest that this component is from small clumps torn from the disc edge
as it truncates, forming a small, variable soft component while the true truncated disc emission is
outside of the bandpass (though it can be seen directly in the lower absorption system XTE J1118+480: 
Esin et al. 2001) As the mass accretion rate
increases, the truncated disc extends closer to the black hole and can be seen directly, though there are
probably still some residual variable clumps which contribute to the spectrum (see also Chiang et al. 2010;
Yamada et al. 2013).

\section{Conclusions}

We present an analysis of the inner disc radius in the low/hard state
of black hole binaries as measured by both disc emission and reflection, 
carefully considering instrumental uncertainties as both features
are typically rather small. 
The limitations of the current calibration are important to consider, as  black hole binaries are extremely bright. This necessitates the
use of the timing (or burst) mode, where the systematics are
not so well understood, yet the excellent quality of the data means that 
the data are typically limited by systematics rather than statistics.

We highlight some calibration issues, such as uncertainties in the 
energy dependence of the off-axis spectral response which become important for piled up spectra
where the central columns are removed. Different energy dependencies 
in the point spread function give grossly different spectra around the 
iron line. We caution that this aspect of 
timing mode is not currently well enough calibrated to constrain
small features such as the smearing of the reflected emission. 
This affects datasets GX4 and especially 
Cyg X-1 in our analysis. Cyg X-1 is the
only low/hard state timing mode observation which requires
an untruncated disc as measured by the reflected spectrum. 

We also discuss the status
of the high energy calibration above $\sim 9$~keV, an uncertainty which
is compounded by the cross-calibration offset of $\Delta \Gamma=0.15$ 
with the {\sl RXTE} PCA in their overlapping bandpass of 3--10~keV, 
which means that they cannot be reliably fit together. A small change in the
effective area above $\sim 9$~keV can dramatically change the 
inferred reflected parameters for strongly smeared reflection. This
affects both S1753 and GX4 in our analysis. For S1753, where 
the standard response shows that reflection is very small but extremely
smeared, a small change in the high energy response (along with a more physically realistic
continuum model) can remove the
requirement for any reflected component. Conversely, for GX4, a
similarly small change can 
dramatically reduce the amount of reflection, switching the solution 
from being reflection dominated to having only a small amount of reflection.
The iron line is strongly smeared in these data, but the models identify this
as being mainly due to ionisation rather than velocity. 

Conversely, observations where reflection is unambiguously detected 
but does not dominate the spectra and is not highly smeared
(i.e. GX1, 2 and 3), the reflection parameters
appear to be fairly robust to these calibration issues. 

The blackbody component is significantly detected in all
datasets irrespective of calibration issues, except in S1753. 
However, in all these data, including S1753, the blackbody is 
independently required by the difference in timing properties at the 
lowest energies (Wilkinson \& Uttley 2009).
We show that the derived inner disc
radius is sensitive to details of the  continuum spectral model, especially in 
S1753 where there is a strong upper limit to the disc radius 
with a single Compton continuum, but where it is much larger with 
the double Compton model which is required 
to produce the continuum spectral lags
(e.g. Wilkinson \& Uttley 2009). 

However, even the double Compton model gives 
inferred inner radii which are somewhat smaller
in the lower luminosity spectra (S1753 and GX1-3) than in GX4
though the radii are still large enough to require the disc to be truncated. 
This could signal a change in the nature of the soft component from 
small variable clumps when the disc is truncated far from the black hole, to 
the truncated disc itself when it extends close enough to the black hole
in the bright low/hard states for this constant emission to contribute to the
low energy bandpass. 

In summary, these EPIC-pn timing mode low/hard state
data do challenge the truncated disc models, but there 
are both model uncertainties and calibration uncertainties which mean that the
challenges can be incorporated by extending the standard, very successful, truncated
disc model rather than abandoning it.  In 
general, we caution that both reflection and the disk blackbody are
small features in this state, so are dependent on the 
details of the instrument calibration as well as on the spectral models
used. We strongly support the on-going effort by the {\sl XMM-Newton} 
team to improve the current calibration and cross-calibration status of
the EPIC-pn.

\section{Acknowledgements}

We would like to thank Matteo Guainazzi for all the extensive
discussions on {\sl XMM-Newton} data reduction and calibration
issues during this project. MK also acknowledges an STFC postdoctoral grant and the support of the Vilho, Yrj\"o and
Kalle V\"ais\"al\"a Foundation from the Finnish Academy of Science. \\ 
This work is based on observations obtained with {\sl XMM-Newton}, an
ESA science mission with instruments and contributions directly funded
by ESA Member States and NASA.


\appendix
\section{The level of calibration uncertainties in the EPIC-pn} 
\label{comparison_crab} 

\subsection{Point spread function}

\begin{figure}
\leavevmode\epsfig{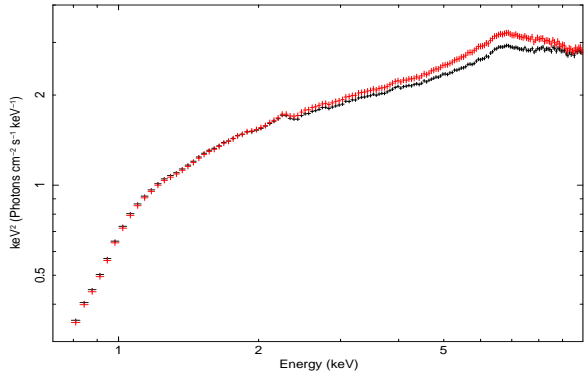}
\caption{GX4 spectra, extracted with SAS v12.0.1. and corrected for pileup by ignoring the central RAWX columns. The black spectrum was extracted using psfmodel=extended, whereas the red spectrum shows the default setting psfmodel=ellbeta. The difference is most pronounced in the iron line region, and the spectral curvature changes above $\sim$7.5~keV. }
\label{ellbeta}
\end{figure}

The SAS v12.0.1. includes an update for the way the point spread function (PSF) is calculated in the {\sc 
arfgen} task. 
The previous `EXTENDED' PSF model is replaced with a new, two-dimensional `ELLBETA' model (Read et al. 
2011) as a default setting. We compared these new responses to the ones generated with SAS v10.0, and  
found that the spectral shape seemed to change noticeably in the two piled-up observations GX4 and Cyg~X-
1 (see Figure~\ref{ellbeta} for the GX4 comparison). Further investigation into this discrepancy showed that 
the `ELLBETA' model indeed has issues at large radii in the timing mode, which become more pronounced 
when the central RAWX columns are excised to correct for pileup. The deviation is most noticeable in the iron 
line region, between $\sim$4--8~keV, and therefore makes a significant difference in any analysis of the 
reflection features. However, manually setting psfmodel=EXTENDED when running {\sc arfgen} brings the 
spectral curvature and the residuals back to a level that is consistent with the previous versions of SAS. 
It is not clear which of these two models better represents the true energy dependence of the PSF. 

\begin{figure}
\begin{center}
\leavevmode\epsfig{file=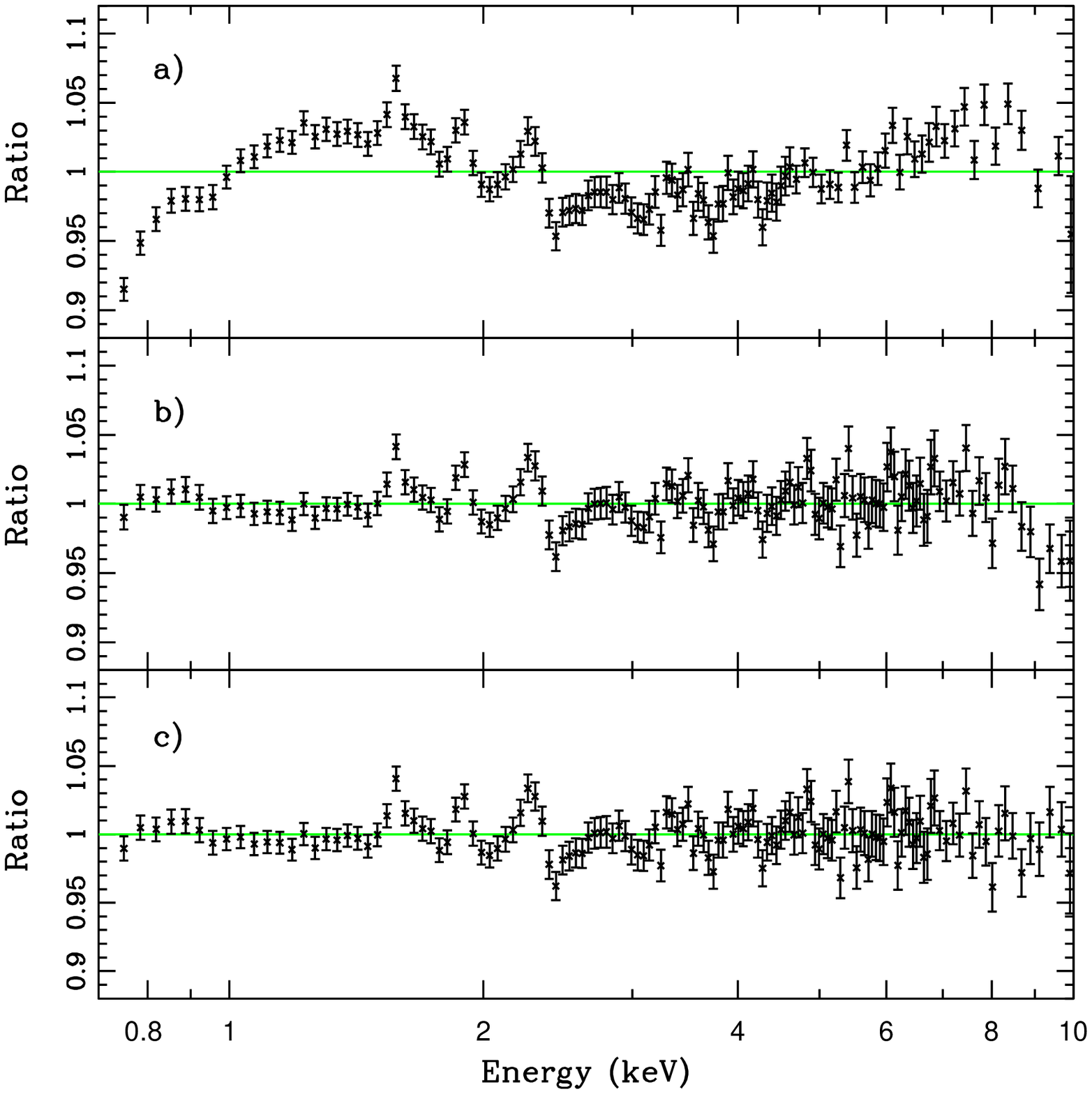,height=8.0cm,width=8cm}
\end{center}
\caption{{\sl Upper panel a):} Residuals from a double power law model of the Crab with {\sc epfast} correction. The features between $\sim$1.6--2.4~keV are due to the CTI correction, which seems to have over-compensated the characteristically negative residuals. {\sl Middle panel b):} A double power law model to account for both nebular and pulsar contributions to the observed spectrum removes the dip at low energies.   {\sl Lower panel c):} Residuals from the same double power law model as the middle panel, with the additional {\sc notch} absorption line fitted at 9.39~keV. } 
\label{crab_resid}
\end{figure}

\subsection{Crab calibration}

We also take a closer look at the drop above 9~keV that is visible in all of our observations using Crab data. 
Figure~\ref{crab_resid} shows a series of residuals from spectral fits to
one of the on-axis observations of the Crab (see Appendix). The top
panel shows residuals to a single power law, with $\Gamma=2.1$. This
is a similar fit to those in Weisskoff et al. 2010, but here the more
appropriate binning does not suppress the visibility of the high
energy residuals. There are clear residuals below 1.5~keV and at 9~keV, which
appear fairly similar to the ones in our data (Fig~\ref{crab_resid}a). However, a single power
law is not a good approximation to the spectrum from the Crab, as
there are both nebular and pulsar contributions. Instead, a double
power law fit removes the soft residuals, but not the high energy
feature at 9~keV (middle panel).  We model this with a {\sc notch},
with width fixed at 1~keV and find a significant reduction in $\chi^2$
(from 260/162 to 231/160) for an energy 9.39$ \pm$0.01~keV, with a
covering fraction of 0.06$\pm$0.02 (i.e. equivalent width 60 eV). The
residuals in Fig~\ref{crab_resid}c are now flat. However, this feature is not
significantly present in the other on-axis Crab spectrum, so we regard this instead as showing the limitations of our
current knowledge of the response.

\subsection{Limits from cross-calibration with RXTE}

One way to assess the status of the cross-calibration between the EPIC-pn fast timing mode and PCA is to 
use Crab data, though this is necessarily in pn burst mode rather than timing mode due to the very high 
count rate of the Crab. Essentially, features seen in the spectra of black hole binaries are not expected to be 
found in the Crab. Thus we tried to shed more light into the very concerning cross-calibrational disagreement 
seen in Figure~\ref{pnpca} by comparing EPIC-pn and PCA observations of the Crab. However, due to the 
brightness of the Crab, most of the {\sl XMM-Newton} observations are taken in a slightly offset position. 
There are only two archived EPIC-pn burst mode observations (01610960401 and 0160960601) that were 
taken in a bore-sight position, thus allowing the whole nebula to be fully encompassed by the aperture (CAL-
TN-0083). 

\begin{figure}
\begin{center}
\leavevmode\epsfig{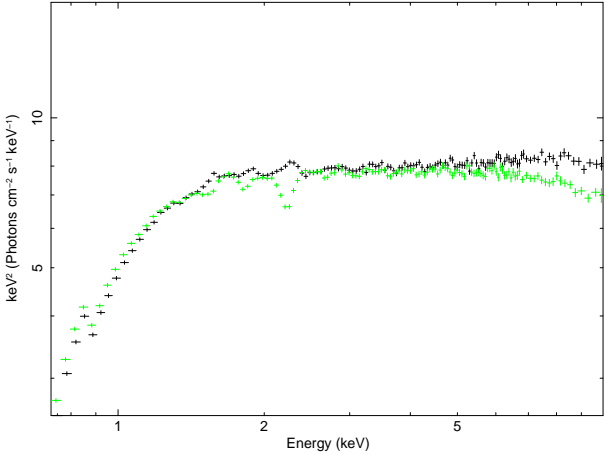}
\end{center}
\caption{Crab observation 01610960401 with {\sc epfast} correction (in black) and without (in green). The rate-dependent CTI effects are visible in the non-corrected data, whereas the corrected one shows over-compensation around the instrumental Si and Au-edges. }
\label{epfast}
\end{figure}

\begin{figure}
\begin{center}
\leavevmode\epsfig{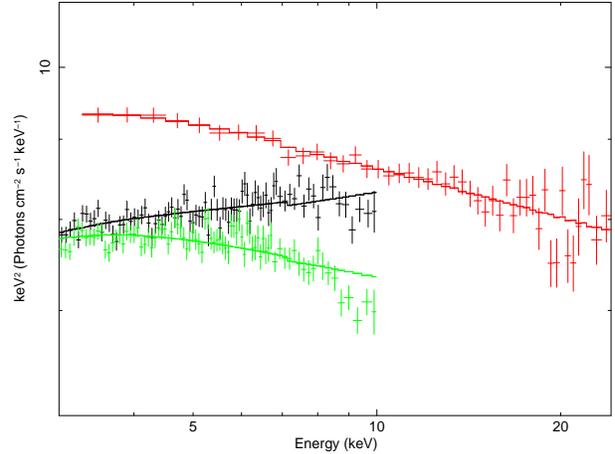}
\end{center}
\caption{Same as Figure~\ref{epfast}, combined with the corresponding PCA observation in red. {\sc epfast} has corrected for the turnover above $\sim$8~keV in the non-corrected data, but rather over-estimated the instrumental edges at $\sim$1.85~keV and $\sim$2.2~keV. }
\label{crab}
\end{figure}

We coupled the pn observation 01610960401 with a PCA observation (90802-02-05-00), taken 5 days prior 
to the {\sl XMM-Newton} one. Since it seems that the rate-dependent CTI correction task {\sc epfast} 
does not properly correct for these pn Crab spectra (see Figure~\ref{epfast} and features around 2~keV in 
Figure~\ref{crab_resid}), we considered both {\sc epfast}-corrected and non-corrected pn spectra for the 
comparison. Figure~\ref{crab} shows the data fitted with an absorbed two-component power law, with the {
\sc epfast}-corrected EPIC-pn data in black, non-corrected in green and the PCA in red. The difference in 
photon index is blatantly evident even by eye, and fitting the spectra in the combined energy range of 0.7--
25~keV gives a difference in photon index of $\Delta\Gamma$=$0.16\pm0.01$ between the CTI-corrected pn 
and the PCA spectra. We tested this further by fitting only above 3~keV to omit any absorption effects that 
might affect the pn spectra at low energies. This yielded individual indices of $1.97^{+0.03}_{-0.06}$ for 
the {\sc epfast}-corrected EPIC-pn, $2.14$ for the non-corrected pn and $2.12$ for the 
PCA. Thus the {\sc epfast}-corrected spectrum shows stronger disagreement with the PCA, whereas the non
-corrected spectrum rolls over above $\sim$8~keV in a way that wrongly resembles the PCA spectral index. 
In addition, the absorption feature seen in Figure~\ref{crab_resid} is even greater in a non-corrected 
spectrum, with a $>$15\% residual at 9~keV.

A discrepancy this noticeable is obviously very concerning. One possibility is that it could be due to an 
incorrect energy-dependency of the rate-dependent CTI and the calibration effort in currently on-going. 
However, dealing with a burst mode observation of the Crab nebula is complex in itself. None of the data 
reduction and/or analysis choices are trivial. Issues such as centralising the extraction region at the peak of 
the emission in RAWX, the size of the extraction region and subtracting a background from the source all 
affect the resulting spectrum. It is also currently impossible to correct for the unknown universal effect of XRL 
in the data. 

However, the issue is not just seen in timing and burst mode spectra. The cross-calibration source for
standard imaging mode, G21.5–0.9, shows the same discrepancy when fit in the 3-10~keV region
of overlap. These data are analysed in the comprehensive cross-calibration paper
of Tsujimoto et al (2011), but he compared datasets in the overlapping 2-10~keV bandpass, so did
not include a high energy PCA--pn comparison.  However, he kindly provided
his data to us, and 
we restrict the bandpass of both to 3-10~keV and find $\Gamma=2.00\pm 0.01$ for
the PCA, with $\Gamma=1.85_{-0.03}^{+0.01}$ for the pn for these imaging mode data.

All these factors in mind, we conclude that the discrepancy in cross-calibration limits joint EPIC-pn/PCA spectral fitting in all modes.

\label{lastpage}

\end{document}